# Topological quantum phase transition in the magnetic semimetal HoSb

J. -M. Zhang[a,*], F. Tang[b,c], Y. -R. Ruan[a,d], Y. Chen[b], R. -W. Zhang[e], W. -T. Guo[a,d], S. -Y. Chen[a,d], J. -P. Li[f], W. Zhao[g], W. Zhou[b], L. Zhang[b], Z. -D. Han[b], B. Qian[b], X. -F. Jiang[b], Z. -G. Huang[a,d], D. Qian[h], and Y. Fang[b,*]

[a]Fujian Provincial Key Laboratory of Quantum Manipulation and New Energy Materials, College of Physics and Energy, Fujian Normal University, Fuzhou 350117, China

[b]Jiangsu Laboratory of Advanced Functional Materials, Department of Physics, Changshu Institute of Technology, Changshu 215500, China

[c]Jiangsu Key Laboratory of Thin Films, School of Physical Science and Technology, Soochow University, Suzhou 215006, China

[d]Fujian Provincial Collaborative Innovation Center for Advanced High-Field Superconducting Materials and Engineering, Fuzhou, 350117, China

[e]Key Lab of advanced optoelectronic quantum architecture and measurement (MOE), Beijing Key Lab of Nanopho tonics &

Ultrafine Optoelectronic Systems, and School of Physics, Beijing Institute of Technology, Beijing 100081, China

[f]Automotive & Transportation Engineering, Shenzhen Polytechnic, Shenzhen 518055, Guangdong, China

[g]ISEM, Innovation Campus, University of Wollongong, Wollongong, New South Wales 2500, Australia

[h]Key Laboratory of Artificial Structures and Quantum Control (Ministry of Education), Shenyang National Laboratory for Materials Science, School of Physics and Astronomy and Tsung-Dao Lee Institute, Shanghai Jiao Tong University, Shanghai 200240, China

**ABSTRACT:** Magnetic topological semimetals, a novel state of quantum matter with nontrivial band topology, have emerged as a new frontier in physics and materials science. An external stimulus like temperature or magnetic field could be expected to alter their spin states and thus the Fermi surface anisotropies and topological features. Here, we perform the angular magnetoresistance measurements and electronic band structure calculations to reveal the evolution of HoSb's Fermi surface anisotropies and topological nature in different magnetic states. The angular magnetoresistance results manifest that its Fermi surface anisotropy is robust in the paramagnetic state but is significantly modulated in the antiferromagnetic and ferromagnetic states. More interestingly, a transition from the trivial (nontrivial) to nontrivial (trivial) topological electronic phase is observed when HoSb undergoes a magnetic transition from the paramagnetic (antiferromagnetic) to antiferromagnetic (ferromagnetic) state induced by temperature (applied magnetic field). Our studying suggests that HoSb provides an archetype platform to study the correlations between magnetism and topological states of matter.

[*)]Author to whom correspondence should be addressed. Electronic mail: jmzhang@fjnu.edu.cn, and fangyong@cslg.edu.cn.

1. **INTRODUCTION**

Rare-earth monoantimonide (*R*Sb) family with a cubic NaCl-type structure have been attracted lots of attentions due to their excellent electronic and magnetic properties.[1-3] Thereinto, the most eye-catching one could be the topological character of accidental crossings between the conduction and valence bands along all high-symmetry lines in the Brillouin zone, which draws a wide range of interest both in the fundamental and applied concerns.[4-7] Nevertheless, a consensus on whether the electronic band structures are indeed topologically nontrivial or not has yet to be formed.[6,8-10] The nonmagnetic *R*Sb compounds with *R* = Sc, Y, La, and Lu have been verified to be trivial semimetals from either experimental or theoretical aspects.[11-15] While, other sister compounds with partially filled *f* shells have long-range magnetic order resulting from the hybridization between the localized 4*f* and conduction electrons,[4,16,17] which brings complexity to the topological properties. The electronic band structures of magnetic *R*Sb compounds will change in different magnetic states driven by either temperature or applied magnetic field. Now a question is that could a topological quantum phase transition, which is a radical departure from the conventional Landau paradigm of symmetry-broken order,[18] take place? An urgent task is to identify the potential candidates and the thermodynamic fingerprints of such exotic properties in the magnetic *R*Sb compounds.

So far, the electronic band structures of magnetic rare-earth monopnictide (*R*Pn) series have been intensively studied.[4,6,8-10,19] While, a common understanding of their topological nature has not yet been achieved. An excellent example is CeSb compound with a magnetic phase diagram including at least 14 distinct spin states. Its topological electronic properties have been hotly debated.[4-6,8-10] Alidoust *et al* observed two sets of temperature-independent Dirac-cone-like dispersion in CeSb by the surface sensitive angle-resolved photoemission spectroscopy (ARPES) using tunable vacuum ultraviolet laser as a photon source and thus speculated the presence of Dirac fermions.[8] Subsequently, Guo *et al* confirmed the band inversion features in HoSb's nonmagnetic (NM) state from theoretical calculations and further revealed the possible emergence of long-sought Weyl fermions in the ferromagnetic (FM) state from both the experimental and theoretical aspects.[20] Kuroda *et al* recently rechecked the electronic band topologies of Ce monopnictides (Pn = P, As, Sb, and Bi) by the bulk-sensitive soft X-ray ARPES and uncovered the topological phase transition from CeP (trivial) to CeBi (nontrivial).[10] Their findings clearly figured out that the topological phase transition significantly depends on the spin-orbit coupling (SOC) strengths in this series and CeSb is actually topologically trivial but closes to the phase transition boundary. It thus can be seen that the electronic properties on the border of a magnetic phase transition are often changeful and their topological nature could be altered in different spin states. This has been proven to be true in several other magnetic $R$Pn members. For instance, Zhou *et al* found that NdSb is topologically trivial in the FM state.[17] While, the chiral anomaly induced negative magnetoresistance observed in the antiferromagnetic (AFM) state suggests that NdSb is a Dirac semimetal.[19] Furthermore, the applied external magnetic field could induce the reconstruction of Fermi surface (FS) following the transition from AFM state to FM state. A similar evolution has also been observed in CeSb compound in our previous work.[4] Nontrivial Berry phase extracted from the de Haas-van Alphen (dHvA) oscillations in CeSb's FM state supports the topological quantum transition from a trivial to a nontrivial electronic state.

As a magnetic $R$Pn, HoSb with a MnO-type AFM ground state shows several magnetic-field-induced spin states.[2] Therefore, its band structure and corresponding topological nature at different magnetic state could be very different. As reported, ARPES measurements found no signature of band inversions above the Néel temperature ($T_N$), which was also confirmed by the theoretical calculations, indicating that HoSb is a topologically trivial semimetal in the paramagnetic (PM) state.[21] On the other hand, nonzero Berry phase deduced from Shubnikov de-Haas (SdH) oscillations in the magnetoresistance implies the quasiparticles in this compound are topologically nontrivial in the FM state[2]. Generally, the band inversions in $R$Pn are deeply below the Fermi level ($E_F$),[4] which bring up a common concern on the reliability of the deduced Berry phase. Thus, alternative approaches either from experimental or theoretical aspects should be adopted to further check the topological nature of the electronic band structures in HoSb. Here, we studied the angular magnetoresistance and electronic band structures at different magnetic states to reveal the temperature and spin structure dependence of FS and the corresponding topological features as well. Our comprehensive observations provide a new insight into the band topology hidden in different magnetic states.

## 2. EXPERIMENTAL AND THEORETICAL METHODS

HoSb crystals were obtained by the flux method using molten tin as a solvent.[2] Magnetization ($M$) studies were performed using a Quantum Design MPMS-SQUID-VSM magnetometer. Specific heat ($C_P$) was measured by a commercial calorimeter in the Physical Property Measurement System (PPMS, Quantum Design) using a thermal relaxation method. Electrical measurements were carried out by the DC resistivity option with sample rotator in PPMS using a standard four-probe method.

Electronic structures were obtained by the density functional theory (DFT) implemented in the Vienna *ab initio* simulation package (VASP).[22,23] The electron-ion interactions were described using the projector augmented wave method and the exchange-correlation part was described by the generalized gradient approximation (GGA) in the scheme of the Perdew-Burke-Ernzerhof (PBE) functional.[24] Here, we employed the lattice constant of $a = b = c = 6.131$ Å for HoSb in our calculation.[25] The plane-wave basis cut-off energy was set to 350 eV in all calculations. And also, a $16 \times 16 \times 16$ Monkhorst $k$-point mesh was adopted. The Ho-4$f$ orbitals in the NM state were regarded as core states. A $2 \times 2 \times 2$ supercell with 16 atoms was constructed to calculate the electronic band structures in the AFM state. On-site Hubbard $U$ for the correlated Ho-4$f$ orbitals (GGA+U) with $U_f = 7$ eV and $J_f = 0.6$ eV were included for calculating the electronic band structures of the magnetic states. In all the calculations, the SOC effect was considered. Moreover, the maximally localized Wannier functions were constructed by employing the WANNIER90 code.[26,27] The corresponding surface states were implemented with the WANNIERTOOLS package based on Greens functions method.[28]

## 3. RESULTS AND DISCUSSIONS

Figure 1(a) shows the magnetization for HoSb as a function of temperature ($T$) along the [001] axis under a magnetic field of 0.01 T. As marked by the arrow, a PM to an AFM phase transition emerges around 5.7 K.[2] The zero-field-cooling (ZFC) and field-cooled cooling (FCC) magnetization curves overlap each other very well, being suggestive of an absence of short-range spin orderings in HoSb. The inset of Fig. 1(a) plots the isothermal magnetization and its derivative $dM/d(\mu_0 H)$ at 2 K, where several field-induced magnetic states including a 1/2 magnetization plateau are observed and their critical fields are indexed. In Fig. 1(b), we present the temperature dependence of specific heat for HoSb. Obviously, a sharp peak generates around $T_N$, confirming an emergence of long-range spin orderings in this compound.[29] Generally, the total specific heat for a certain material with spontaneous spin orderings can be expressed as $C_p = C_m + C_e + C_l$, where $C_m$, $C_e$, and $C_l$ represent the magnetic, electronic, and lattice contributions, respectively.[30] Here, as shown in Fig. 1(b), the structurally similar, LaSb is employed as a reference material to approach the $C_e + C_l$ of HoSb. Thus, $C_m/T$ and the magnetic entropy $S_m$ for HoSb can be obtained. It can be found that in the inset of Fig. 1(b) $S_m$ saturates to 14 J·K$^{-1}$·mol$^{-1}$, which is only about 75% of the expected value (18.6) of $R\ln(2J+1)$ with $J = 8$ for Ho$^{3+}$,[31] suggesting that the thermal population of crystal-field-split Ho$^{3+}$ states is not fully complete at this temperature.[30] Figure 1(c) plots the temperature-dependent resistivity ($\rho$) for HoSb under zero field from 300 to 2 K. As shown, the zero-field resistivity decreases with the decreasing temperature, exhibiting metallic property over the whole temperature range. Above 85 K, the linear temperature dependence in resistivity suggests that electron-phonon scattering dominates the transport.[30] With the temperature further decreasing, the resistivity significantly deviates from linear temperature dependence and shows a sharp drop around 6 K, which could be ascribed to the suppressed spin-disorder scattering at the magnetically ordered state.[32] Below this temperature, the zero-field resistivity does not simply follow a quadratic temperature ($T^2$) dependence [see upper-left inset of Fig. 1(c)], indicating the possible absence of Fermi-liquid states as revealed in some other semimetals like WTe$_2$, PtBi$_2$, and so on.[33,34] Upon application of magnetic fields, the resistivity at low temperature is significantly enhanced. As plotted in the lower-right inset of Fig. 1(c), the magnetoresistance (*MR*) reaches 3.2×10$^4$% at 2 K and 9 T. Different to the generally observed $B^2$ dependence in the magnetoresistance for those NM sister compounds $R$Pn (Pn = Sb and Bi), an inflexion around 4 T is found, which can be related to the field-induced magnetic state transition, suggesting the cross coupling between spin order and electrical properties. Assuming that the hole-to-electron concentration ratio in HoSb equals to 1, a rough fitting of the magnetoresistance at 2 K with $MR = (u_{ave} \times B)^2$ yields $u_{ave}$ ~ 2.23×10$^4$ cm$^2$·V$^{-1}$·s$^{-1}$,[30] which is comparable to those obtained in other $R$Pn (Pn = Sb and Bi).[4,32] To better understand the transport behavior in HoSb, we perform the magnetoresistance measurements under different temperatures (not shown) and evaluate Kohler's rule scaling using $MR \sim F(B/\rho_0)$ that takes effect when there is only a single scattering rate or several scattering rates with unchanged relative contribution are just as true.[30] Here, $\rho_0$ is zero-field resistivity at a given temperature. Figure 1(d) shows the Kohler plot for HoSb. The curves cover two temperature regions: the first region is below 10 K and the second one is above 50 K. Clearly, the high-temperature magnetoresistance curves well obey the Kohler's rule, indicating that a unique temperature-dependent scattering relaxation time for the carriers and single band approximation is sufficient to describe the transport process for HoSb.[30] While, the violation of Kohler's rule at low temperature could be derived from the multiband effect or multiple scattering processes, which has been confirmed by the Hall resistivity in ref. 2.

To trace the FS evolution for HoSb, we perform the magnetoresistance measurement at different angles and several selected temperatures. Figures 2(a)-2(c) depict the isothermal magnetoresistance at fixed angles ($\theta$) and the temperatures above 10 K (> $T_N$). A sketch of our experimental setup is plotted in the inset of Fig. 2(a), where the current flows along the $a$ axis and the magnetic field is rotated in the $ac$ plane. Here, the magnetic field angle is measured from the [001] axis to the [100] axis. As shown, the magnetoresistance in Figs. 2(a)-2(c) decreases with the angles increasing from 0° to 90°. This implies that the magnetoresistance is governed by the normal component of applied magnetic field ($\mu_0 H|\cos\theta|$),[30,32] which is confirmed by the angular magnetoresistance plotted in Fig. S1(a). The arresting field dependence of magnetoresistance at 90° contradicts with the case as expected in a 2D system where the resistance should be field-independent since only the $\mu_0 H|\cos\theta|$ contributes to the magnetoresistance.[30] Generally, in light of the simple Drude free-electron model, the resistivity should be directly related to the charge carrier mobility $\mu$, namely $\rho = ne\mu$ with $\mu = e\tau/m$, where $e$ is the electron charge, $\tau$ is the relaxation time, and $m$ is the effective mass.[35] It thus can be seen that the angular magnetoresistance is expected to be regulated by the effective mass anisotropy. Typically, for an elliptical-shaped FS, its anisotropic quantity $Q(H, \theta)$ obeys such a scaling relation, $Q(H, \theta) = Q(\varepsilon_\theta H)$, which is the case in WTe$_2$, high-$T_c$ superconductor and so on.[35,36] Here, $\varepsilon_\theta H$ is the reduced magnetic field and $\varepsilon_\theta = (\cos^2\theta + \gamma^{-2}\sin^2\theta)^{1/2}$ is a signature of the mass anisotropy with $\gamma^2$ being the ratio of the effective masses of electrons moving in the direction from 0° to 90°. As referred in our early work on ErBi[30] and ref. 37, the transport properties of $R$Pn series are dominated by the electron pockets and the elliptical FS governs the angular magnetoresistance. Thus, in HoSb, the temperature and spin effect on its band structures could also be reflected by the angular-dependent isothermal magnetoresistance. As shown in Figs. 2(d)-2(f), the magnetoresistance isotherms measured at different angles in Figs. 2(a)-2(c) collapse onto a single master line, namely the $MR$ curve at 0°, after scaling by employing $\varepsilon_\theta = (\cos^2\theta + \gamma^{-2}\sin^2\theta)^{1/2}$, where $\gamma$ is a constant at a given temperature. The obtained $\varepsilon_\theta$ at different temperatures are plotted in Fig. 2(g). Clearly, a strong angle dependence is revealed, implying the 3D FS nature in HoSb. Fitting the variation of $\varepsilon_\theta$ at 150 K with respect of $\theta$, $\gamma$ is assigned to be 6.9, which closes to as obtained in LaBi (~ 7.9) but larger than that of WTe$_2$ (~ 2).[35] After performing the same operation on the magnetoresistance isotherms at 100 K and 10 K, it can be found that the obtained $\varepsilon_\theta$ barely changes, suggesting that the FS anisotropy of HoSb is robust against temperature.

Figures 3(a)-3(c) show the field-dependent magnetoresistance for HoSb at various angles below $T_N$. Different to the parabolic-field-dependent magnetoresistance in those NM $R$Pn,[1,4] the low-temperature magnetoresistance for HoSb are significantly affected by the spin orderings, which can also be reflected by the angular magnetoresistance shown in Fig. S1, indicating that the electronic band structure could be altered in the ordered magnetic states. Besides, the critical fields for magnetoresistance anomalies (corresponding to the magnetic phase transitions) in HoSb vary from one angle to another, revealing their complicated angle dependences. As reported, due to the strong $p$-$f$ mixing, the magnetization easy axis for HoSb is in the [001] direction.[2] Thus, the magnetic transitions between different spin states are determined by the magnetic field component along the magnetization easy axis and their critical magnetic fields should have a 1/cos$\varphi$ dependence, where $\varphi$ is the angle between the magnetization and the [001] axis. Note that if the magnetization easy axis is always fixed in the [001] axis, the expected critical transitions fields should consecutively increase from 0° to 90° in a monotonous way. As shown in Fig. 3(c), this is indeed the case at the angles less than 20°. While, with the angles further increasing, especially above 60°, the critical phase transition field monotonously decreases. It is out of expectation, but is within understanding due to the cubic crystal symmetry for HoSb. This is reminiscent of the field-induced magnetic phase transition in CeSb, where the $\Gamma_8$ orbital state flops from [001] axis to [010] or [100] axis at the angles of $(2n+1)\pi/4$ ($n$=0, 1, 2, and 3).[38,39] Meanwhile, the magnetization

easy axis changes its direction in the same way as well. Consequently, the critical phase transition fields for CeSb repeat their values every 90°.[38] Similar case could be expected in HoSb, since in a cubic crystal the *a*, *b* and *c* axes are equivalent. Considering the complex angle dependence of critical phase transition fields and possible rearrangement of the magnetization easy axis, the FS anisotropies of HoSb at low temperatures could not be simply understood by the above-mentioned method and thus an alternative way should be employed.

From the magnetoresistance data shown in Figs. 2 and 3, it is clear that both the temperature and magnetic field contribute to the FS anisotropy of HoSb. We perform the DFT calculations to further reveal how the two ingredients influence the electronic properties. Figures 4(a)-4(c) show HoSb's spin configuration at three different magnetic states including the NM, AFM and FM phases respectively, and the corresponding bulk Brillouin zone of this compound [see Fig. 4(d)]. Next, we focus on the atomic orbital composition of HoSb at the NM, FM and AFM states. Figure 4(e) plots the partial density of states (PDOS) for HoSb at the NM state. The hybrid Ho-*d* and Sb-*p* orbitals contribute almost equally to the occupied state and the non-occupied state is mainly from the Ho-*d* states. Note that the low DOS near the $E_F$ confirms that HoSb is a semimetal, which has been a consensus on the electronic properties of the *R*Pn (Pn=Sb and Bi) series.[30,32] At the AFM and FM states, as shown in Figs. 4(f)-4(g), the Ho-*d* and Sb-*p* orbital distributions are not significantly altered, while the Ho-4*f* electrons are highly localized in the non-occupied states located between ~ 0.5 and 1.5 eV and -2.0 to -1.5 eV, respectively. Note that the PDOS in the AFM state displays a small broad hump at the Fermi level, indicating that the Ho-*d* and Sb-*p* states host more localized features respect to those in the other two distinct magnetic phases.

To further understand the electronic properties of HoSb, the projected band structures of Ho-*d*, Ho-*f,* and Sb-*p* states are calculated. Here, the generalized gradient approximation (GGA) plus SOC is adopted for the electron part. As plotted in Fig. 5 (a), the bulk band structures at the NM state are double degenerate. Band inversions happen at ∼ 0.7 eV above and below $E_F$ near the Γ and X points, respectively. There is a fairly small band gap around the X point between the conduction and valence bands. Using parity analysis,[40] the $Z_2$ topological invariant is evaluated to be (1; 000), which is confirmed by the calculated Wilson loop (see Fig. S2 in the Supplementary Information), suggesting that HoSb host topologically nontrivial electronic properties at the NM state. However, generally, in the ordinary GGA calculations, the SOC effect is overestimated, which results in the underestimated energy gaps. Therefore, more accurate band structure calculations using the HSE06 functional is performed on HoSb to check its topological nature.[41] Comparing to GGA results, as shown in Fig. 5(b), much larger band gaps at the Γ and X points, and the degenerate valence band splitting at X point are obtained in HSE06 calculations. No band inversions emerge in HSE06 band structures, indicating that HoSb is a topologically trivial semimetal in the NM state. Fig. 5(c) plots HoSb's FSs with three double degenerate Fermi pockets at its NM phase. Two hole pockets (β and γ) center at the Brillouin zone and triplicate electron pockets ($α_1$, $α_2$, and $α_3$) elongate along the Γ-X direction.

As mentioned above, the ordered magnetic structures in HoSb could alter its FS anisotropy. Thus, it is necessary to check the difference of the electronic band structures between different magnetic states. Here, the band structures of this compound at its AFM state are evaluated. Figure 6(a) shows the electronic band structure of AFM HoSb from the GGA calculation. Note that due to the increased atom numbers adopted in the band calculations, the band numbers near the $E_F$ increase as well, together with a certain degree of band folding. As plotted in Fig. 6(b), the two opposite spins cancel each other out and thus lead to zero net magnetic moment. As a consequence, the band structures remain double-degenerate in the AFM state of HoSb. Obviously, the band anti-crossings, where the Sb-$p$ and Ho-$d$ states are inverted, emerges near the Γ-X path, indicating that the AFM HoSb is a topologically nontrivial semimetal. Here, the band gaps around the inverted bands in the AFM state has been optimized by the hybrid functional HSE06 calculations. Clearly, as observed that in the GGA calculations, the band inversion [shown in Fig. 6(b)] exists even though a gap of ~ 0.1 eV is opened in the band structures, suggesting that the topological feature of the electronic band structures always holds true. Note that comparing to the results obtained by the GGA calculations the HSE06 band structures are less densely sampled for simplicity. Even so, the topological electronic properties of AFM HoSb can still be easily extracted. Figure 6(c) depicts HoSb's FS consisting of one hole pocket centered at the Γ point and one electron pocket with one part wrapped at the Γ point and the other part located on the Γ-X path in triplicate at the AFM state. Figures 6(d)-6(e) plot the calculated Fermi pockets projection in (001) and (110) planes. It can be seen that the electron pocket and hole pocket extend to each other, which clearly shows the difference to those observed in the Fermi pockets in HoSb's NM state.

Figure 6(f) displays the surface spectrum of the (111) plane for the AFM state. As revealed in Figs. 6(f)-6(g), obvious band crossovers emerge at ~ -0.1 eV between $\bar{\Gamma}$ - $\bar{M}$. For clarity, the iso-energy surfaces [see Fig. 6(g)] enclosing the $\bar{\Gamma}$ point at the Fermi level for the (111) plane are extracted, which is basically consistent with the case of Fig. 6(c). Note that the section morphology of 'snowflake' shaped ring conclusively shows the topologically non-trivial nature of the AFM state in HoSb.

Recently, clear magnetoresistance oscillations were observed in the FM state of HoSb,[2] from which the nontrivial Berry phase was extracted. Thus, one could expect the emergence of Weyl fermions in HoSb's FM state. Here, we calculated the electronic band structures in the FM state of HoSb *via* both the GGA and HSE06 methods. As shown in Fig. 7(a), the GGA calculations suggest that due to the breaking time-reversal symmetry in the field-induced NM state the double degeneracy along the Γ-X path is split. Note that the band inversions still hold and a pair of band crossings with very small band gaps [shown in the inset of Fig. 7(a)] occur at point X. Such a case is quite similar to the twofold-degenerate Weyl points observed in CeSb.[4] As referred above, the GGA calculations always raise an underestimate of the band gaps. Thus, we further check the electronic band structures of the FM state in HoSb by the HSE06 calculations. It can be found that in Fig. 7(b) the band structures obtained using the latter approach are significantly altered. As shown, the band gaps, of which only the amplitudes increase, still exist at the Γ and X points, while the band inversions disappear. It thus can be seen that similar to that obtained in the NM state, the electronic properties of the FM state in HoSb are topologically trivial. Figure 7(c) shows the bulk FS of FM HoSb, in which two sets of electron pockets centered at X and elongated along the Γ-X path and two hole pockets centered at Γ point are observed. Note that in the fully polarized magnetic states the spin degeneracy is released and thus degenerate FS pockets are split due to the field-induced time-inversion symmetry breaking.

## 4. CONCLUSION

We report the isothermal magnetoresistance at different fixed angles and various temperatures and the electronic band structures in different spin states for HoSb single crystals. The magnetoresistance at high temperature ($>T_N$) show a parabolic field dependence as observed in other NM sister compounds. While, the field-induced magnetoresistance anomalies are observed at low temperature ($<T_N$) and their critical fields change the values from one angle to another, reflecting the spin effect on the angle-dependent magnetoresistance. HoSb's FS topology has a strong spin ordering dependence, which is further confirmed by our theoretical calculations. Besides, the topological properties for the three magnetic states of HoSb are discussed. These findings provide a paradigm for the field effect on control of topological quantum phase transition and are relevant to the understanding of rare-earth-based magnetic topological materials.

**Author contributions**

J.-M. Zhang and Y. Fang conceived and supervised this work. F. Tang, Y. Chen, W. Zhou, L. Zhang, S. -Y. Chen, J.-P. Li and W. Zhao prepared and analyzed the samples, Y.-R. Ruan, W.-T. Guo and R.-W. Zhang performed theoretical calculation. J.-M. Zhang, Y. Fang, F. Tang and Y.-R. Ruan wrote the manuscript. Z.-D. Han, B. Qian, X.-F. Jiang, Z.-G. Huang and D. Qian analyzed the datas and provided valuable and constructive suggestions.

**Conflicts of interest**

The authors declare no competing financial interests.

**Acknowledgements**

This work is supported by the Key University Science Research Project of Jiangsu Province (No. 19KJA530003), National Natural Science Foundation of China

(Nos. 11604027, 11874113, 11704047, U1832147 and 12047512), Natural Science Foundation of Fujian Province of China (No. 2020J02018), Natural Science Foundation of Guangdong Province (No. 2017A030310578), Open Fund of Fujian Provincial Key Laboratory of Quantum Manipulation and New Energy Materials (No. QMNEM1903), and the Project Funded by China Postdoctoral Science Foundation (No.2020M680011).

**References**

(1) F. F. Tafti, Q. D. Gibson, S. K. Kushwaha, N. Haldolaarachchige and R. J. Cava, *Nat. Phys*., 2015, **12**, 272.

(2) Y. Y. Wang, L. L. Sun, S. Xu, Y. Su and T. L. Xia, *Phys. Rev. B*, 2018, **98**, 045137.

(3) P. G. LaBarre, B. Kuthanazhi, C. Abel, P. C. Canfield and A. P. Ramirez, *Phys. Rev. Lett.*, 2020, **125**, 247203.

(4) Y. Fang, F. Tang, Y. R. Ruan, J. M. Zhang, H. Zhang, H. Gu, W. Y. Zhao, Z. D. Han, W. Tian, B. Qian, X. F. Jiang, X. M. Zhang and X. Ke, *Phys. Rev. B*, 2020, **101**, 094424.

(5) K. Kuroda, Y. Arai, N. Rezaei, S. Kunisada, S. Sakuragi, M. Alaei, Y. Kinoshita, C. Bareille, R. Noguchi, M. Nakayama, S. Akebi, M. Sakano, K. Kawaguchi, M. Arita, S. Ideta, K. Tanaka, H. Kitazawa, K. Okazaki, M. Tokunaga, Y. Haga, S. Shin, H. S. Suzuki, R. Arita and T. Kondo, *Nat. Commun.*, 2020, **11**, 2888.

(6) Y. Wu, Y. Lee, T. Kong, D. Mou, R. Jiang, L. Huang, S. L. Bud'ko, P. C. Canfield and A. Kaminski, *Phys. Rev. B*, 2017, **96**, 035134.

(7) S. Chatterjee, S. Khalid, H. S. Inbar, A. Goswami, F. C. de Lima, A. Sharan, F. P. Sabino, T. L. Brown-Heft, Y. H. Chang, A. V. Fedorov, D. Read, A. Janotti and C. J. Palmstrøm, *Phys. Rev. B*, 2019, **99**, 125134.

(8) N. Alidoust, A. Alexandradinata, S. Y. Xu, I. Belopolski, S. K. Kushwaha, M. Zeng, M. Neupane, G. Bian, C. Liu, D. S. Sanchez, P. P. Shibayev, H. Zheng, L. Fu, A. Bansil, H. Lin, R. J. Cava and M. Z. Hasan, *arXiv:1604.0857*, 2016.

(9) H. Oinuma, S. Souma, D. Takane, T. Nakamura, K. Nakayama, T. Mitsuhashi, K. Horiba, H. Kumigashira, M. Yoshida, A. Ochiai, T. Takahashi and T. Sato, *Phys. Rev. B*, 2017, **96**, 041120(R).

(10) K. Kuroda, M. Ochi, H. S. Suzuki, M. Hirayama, M. Nakayama, R. Noguchi, C. Bareille, S. Akebi, S. Kunisada, T. Muro, M. D. Watson, H. Kitazawa, Y. Haga, T. K. Kim, M. Hoesch, S. Shin, R. Arita and T. Kondo, *Phys. Rev. Lett.*, 2018, **120**, 086402.

(11) Y. J. Hu, E. I. Paredes Aulestia, K. F. Tse, C. N. Kuo, J. Y. Zhu, C. S. Lue, K. T. Lai and Swee K. Goh, *Phys. Rev. B*, 2018, **98**, 035133.

(12) J. He, C. Zhang, N. J. Ghimire, T. Liang, C. Jia, J. Jiang, S. Tang, S. Chen, Y. He, S. K. Mo, C. C. Hwang, M. Hashimoto, D. H. Lu, B. Moritz, T. P. Devereaux, Y. L. Chen, J. F. Mitchell and Z. X. Shen, *Phys. Rev. Lett.*, 2016, **117**, 267201.

(13) L. K. Zeng, R. Lou, D. S. Wu, Q. N. Xu, P. J. Guo, L. Y. Kong, Y. G. Zhong, J. Z. Ma, B. B. Fu, P. Richard, P. Wang, G. T. Liu, L. Lu, Y. B. Huang, C. S. S. Fang, Sun, Q. Wang, L. Wang, Y. G. Shi, H. M. Weng, H. C. Lei, K. Liu, S. C. Wang, T. Qian, J. L. Luo and H. Ding, *Phys. Rev. Lett.*, 2016, **117**, 127204.

(14) S. Khalid, A. Sharan and A. Janotti, *Phys. Rev. B*, 2020, **101**, 125105.

(15) O. Pavlosiuk, M. Kleinert, P. Swatek, D. Kaczorowski and P. Wiśniewski, *Sci. Rep.*, 2017, **7**, 12822.

(16) W. Liu, D. Liang, F. Meng, J. Zhao, W. Zhu, J. Fan, L. Pi, C. Zhang, L. Zhang and Y. Zhang, *Phys. Rev. B*, 2020, **102**, 174417.

(17) Y. Zhou, X. Zhu, S. Huang, X. Chen, Y. Zhou, C. An, B. Zhang Y., Yuan, Z. Xia, C. Gu and Z. Yang, *Phys. Rev. B*, 2017, **96**, 205122.

(18) N. L. Nair, P. T. Dumitrescu, S. Channa, S. M. Griffin, J. B. Neaton, A. C. Potter and J. G. Analytis, *Phys. Rev. B*, 2018, **97**, 041111(R).

(19) Y. Wang; J. H. Yu, Y. Q. Wang, C. Y. Xi, L. S. Ling, S. L. Zhang, J. R. Wang, Y. M. Xiong, T. Han, H. Han, J. Yang, J. Gong, L. Luo, W. Tong, L. Zhang, Z. Qu, Y. Y. Han, W. K. Zhu, L. Pi, X. G. Wan, C. Zhang and Y. Zhang, *Phys. Rev. B*, 2018, **97**, 115133.

(20) C. Y. Guo, C. Cao, M. Smidman, F. Wu, Y. J. Zhang, F. Steglich, F. C. Zhang and H. Q. Yuan, *npj Quantum Mater.*, 2017, **2**, 39.

(21) M. M. Hosen, G. Dhakal, B. Wang, N. Poudel, B. Singh, K. Dimitri, F. Kabir, C. Sims, S. Regmi, W. Neff, A. B. Sarkar, A. Agarwal, D. Murray, F. Weickert, K. Gofryk, O. Pavlosiuk, P. Wiśniewski, D. Kaczorowski, A. Bansil and M. Neupane, *Sci. Rep.*, 2020, **10**, 12961.

(22) G. Kresse and J. Hafner, *Phys. Rev. B*, 1993, **48**, 13115.

(23) G. Kresse and J. Furthmüller, *Phys. Rev. B*, 1996, **54**, 11169.

(24) J. P. Perdew, K. Burke and M. Ernzerhof, *Phys. Rev. Lett.*, 1996, **77**, 3865.

(25) M. N. Abdusaljamova, O. R. Burnashev and K. E. Mironov, *J. Less-Common Met.*, 1984, **102**, L19.

(26) N. Marzari and D. Vanderbilt, *Phys. Rev. B*, 1997, **56**, 12847.

(27) I. Souza, N. Marzari and D. Vanderbilt, *Phys. Rev. B*, 2001, **65**, 035109.

(28) Q. Wu, S. Zhang, H. F. Song, M. Troyer and A. A. Soluyanov, *Comput. Phys. Commun.*, 2018, **224**, 405.

(29) T. O. Brun, F. W. Korty and J. S. Kouvel, *J. Magn. Magn. Mater.*, 1980, **15**, 298.

(30) L. Y. Fan, F. Tang, W. Z. Meng, W. Zhao, L. Zhang, Z. D. Han, B. Qian, X. F. Jiang, X. M. Zhang and Y. Fang, *Phys. Rev. B*, 2020, **102**, 104417.

(31) D. Neogy, K. N. Chattopadhyay, P. K. Chakrabarti, H. Sen and B. M. Wanklyn, *J. Magn. Magn. Mater.*, 1996, **154**, 127.

(32) J. J. Song, F. Tang, W. Zhou, Y. Fang, H. L. Yu, Z. D. Han, B. Qian, X. F. Jiang, D. H. Wang and Y. W. Du, *J. Mater. Chem. C*, 2018, **6**, 3026.

(33) Y. L. Wang, L. R. Thoutam, J. Z. Xiao, L. Hu, S. Das, Z. Q. Mao, J. Wei, R. Divan, A. Luican-Mayer, G. W. Crabtree and W. K. Kwok, *Phys. Rev. B*, 2015, **92**, 180402(R).

(34) W. Gao, N. Hao, F. W. Zheng, W. Ning, M. Wu, X. Zhu, G. Zheng, J. Zhang, J. Lu, H. Zhang, C. Xi, J. Yang, H. Du, P. Zhang, Y. Zhang and M. Tian, *Phys. Rev. Lett.*, 2017, **118**, 256601.

(35) L. R. Thoutam, Y. L. Wang, Z. L. Xiao, S. Das, A. Luican-Mayer, R. Divan, G. W. Crabtree and W. K. Kwok, *Phys. Rev. Lett.*, 2015, **115**, 046602.

(36) J. Du, Z. Lou, S. Zhang, Y. Zhou, B. Xu, Q. Chen, Y. Tang, S. Chen, H. Chen, Q. Zhu, H. Wang, J. Yang, Q. Wu, Oleg V. Yazyev and M. Fang, *Phys. Rev. B*, 2018, **97**, 245101.

(37) B. Qian, F. Tang, Y. R. Ruan, Y. Fang, Z. D. Han, X. F. Jiang, J. M. Zhang, S. Y. Chen and D. H. Wang, *J. Mater. Chem. C*, 2018, **6**, 10020.

(38) J. Xu, F. Wu, J. K. Bao, F. Han, Z. L. Xiao, I. Martin, Y. Y. Lyu, Y. L. Wang, D. Y. Chung, M. Li, W. Zhang, J. E. Pearson, J. S. Jiang, M. G. Kanatzidis and W. K. Kwok, *Nat. Commun.*, 2019, **10**, 2875.

(39) Linda Ye, Takehito Suzuki, Christina R. Wicker and Joseph G. Checkelsky, *Phys. Rev. B*, 2018, **97**, 081108(R).

(40) L. Fu and C. L. Kane, *Phys. Rev. B*, 2007, **76**, 045302.

(41) J. Heyd, G. E. Scuseria and M. Ernzerhof, *J. Chem. Phys.*, 2003, **118**, 8207.

**Figures and captions**

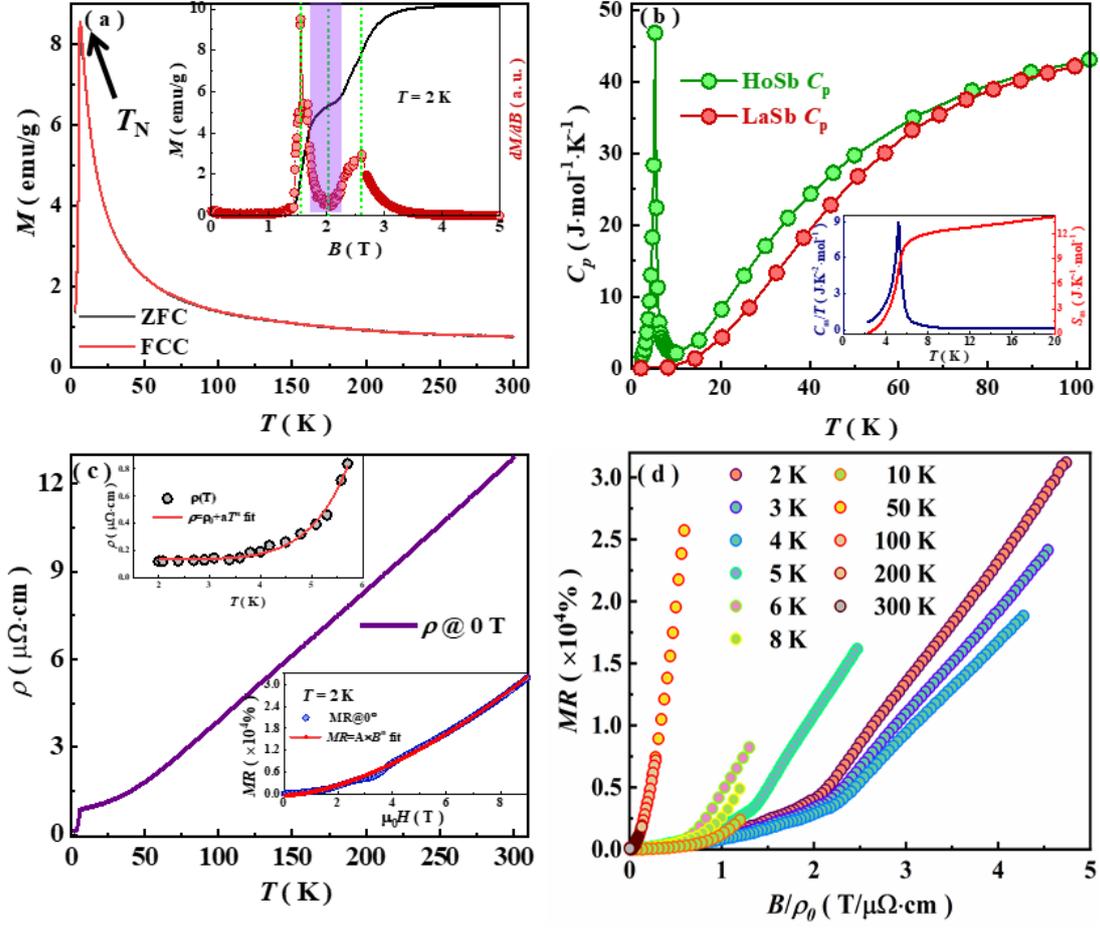

**Fig. 1** (a) ZFC and FCC magnetization as a function of temperature for HoSb. The inset shows the magnetization and isothermal magnetization and its derivative $dM/d(\mu_0 H)$ at 2 K. (b) Temperature-dependent specific heat for HoSb and LaSb. The inset plots the magnetic specific heat divided by temperature and magnetic entropy for HoSb as a function of temperature. (c) The temperature-dependent zero-field resistivity for HoSb. The left upper inset shows the low-temperature part of resistivity and its fit using $\rho = \rho_0 + aT^n$. (d) Kohler's plot for the magnetoresistance in HoSb.

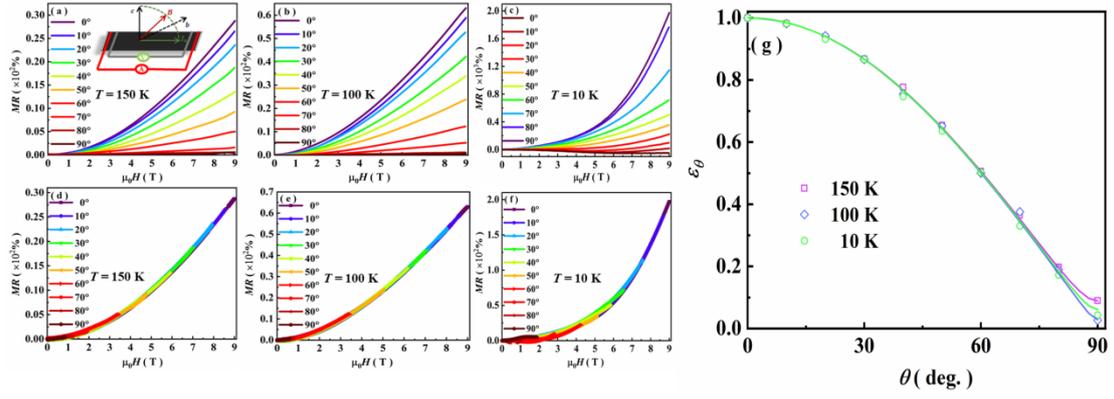

**Fig. 2** (a)-(c) Magnetoresistance as a function of magnetic field at different angles above $T_N$. The inset of (a) shows the definition of measured magnetic field. (d)-(f) Data in (a)-(c) replotted with $\mu_0 H$ scaled by a factor $\varepsilon_\theta$.

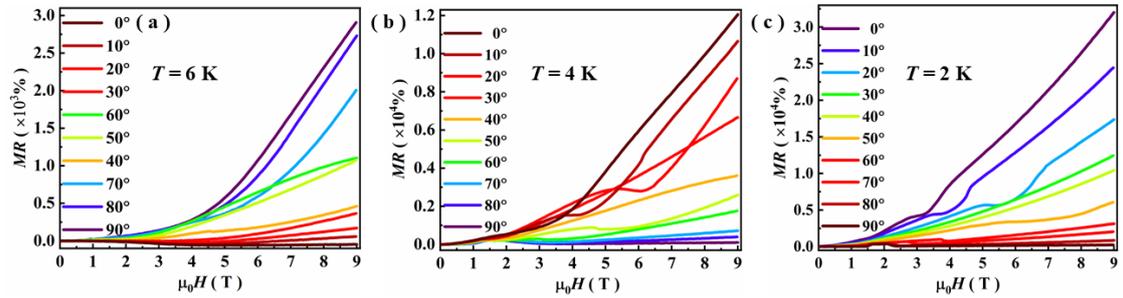

**Fig. 3** Magnetic field dependence of magnetoresistance at (a) 6 K, (b) 4 K, and (c) 2 K.

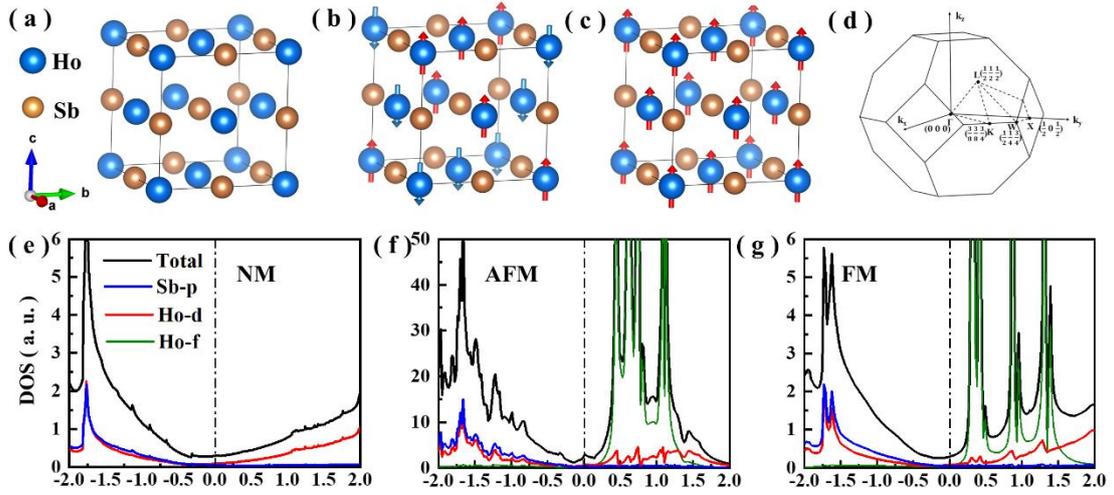

**Fig. 4** Spin configurations of HoSb at the (a) NM state, (b) AFM state, and (c) FM state. (d) High-symmetry points in the Brillouin zone of HoSb. (e)-(g) PDOS of HoSb in the NM, AFM and FM states, respectively. The dotted line denotes the $E_F$.

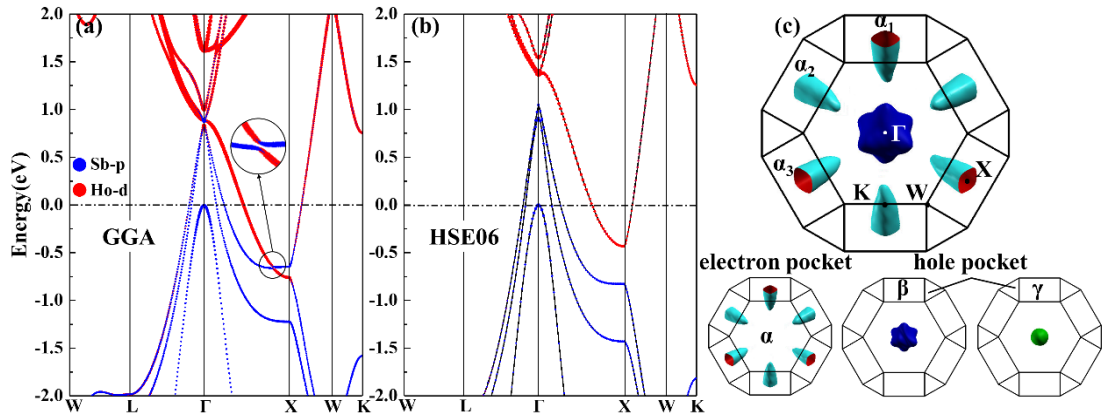

**Fig. 5** Projected band structure of HoSb in the NM states obtained using the (a) GGA and (b) HSE06 methods. A small gap can be found near the X point of Fig. 5(a). The red and blue dots denote the Ho-$d$ and Sb-$p$ orbital contribution, respectively. The dotted line denotes the $E_F$. (c) HoSb' FS in the NM states, which consists of one electron pocket (α) and two hole pockets (β and γ).

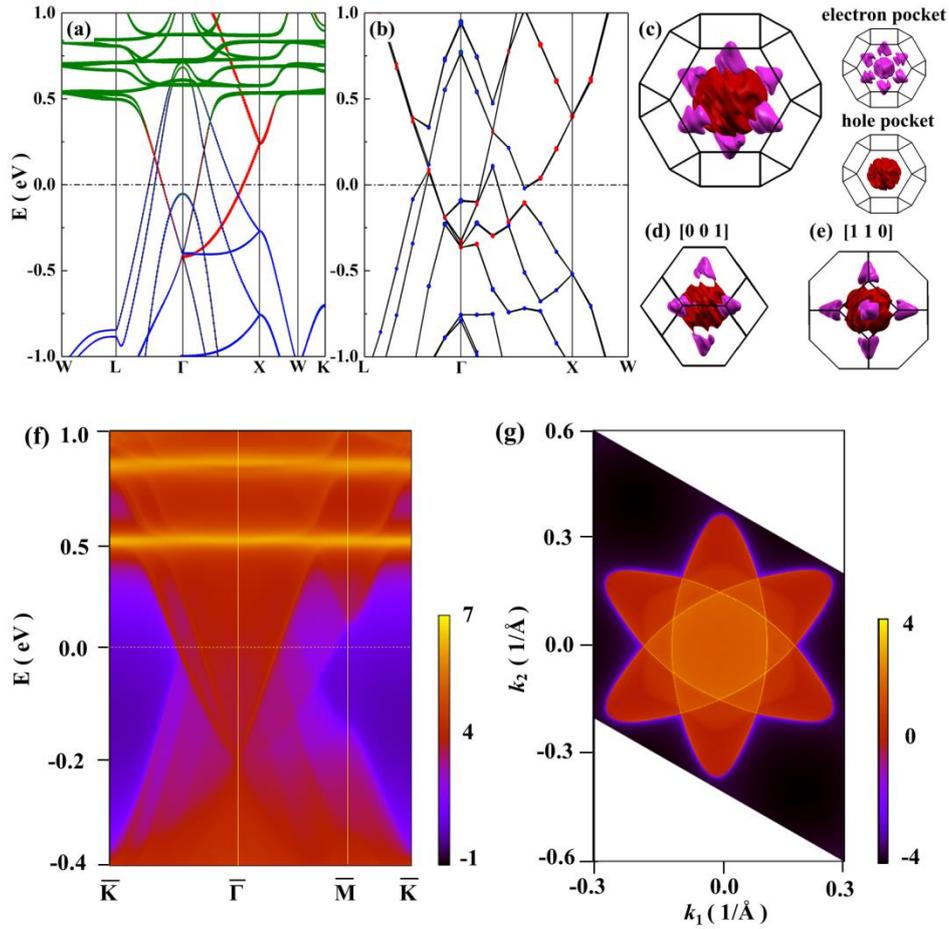

**Fig. 6** Projected band structure of HoSb in the AFM state calculated using the (a) GGA and (b) HSE06 method. The red and blue dots denote the Ho-*d* and Sb-*p* orbital contributions, respectively. The dotted line denotes the $E_F$. (c) HoSb's FS in the AFM state, and its views from (d) (001) and (e) (110) planes. Here, the FS consists of one electron pocket and one hole pocket. (f) The projected surface states on the (111) plane for HoSb's AFM state. (g) The iso-energy ($E_F$) surface around the $\bar{\Gamma}$ point on the (111) plane.

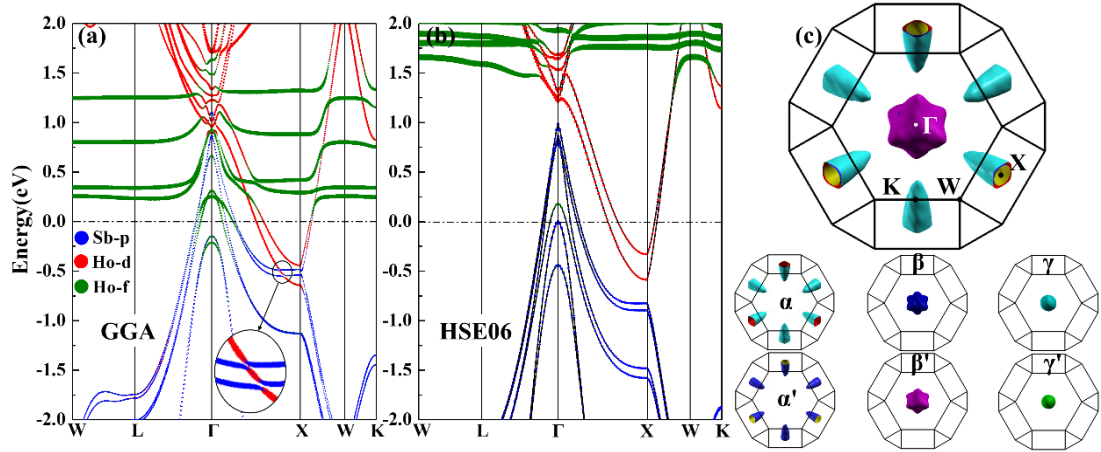

**Fig. 7** Projected band structure of HoSb in the FM state obtained using the (a) GGA and (b) HSE06 method. Two small gaps are observed near X point in Fig. 7(a). The red dots and blue dots represent the Ho-*d* and Sb-*p* orbital contribution, respectively. The dotted line denotes the $E_F$. (c) HoSb's FS in the FM state.

## Supplementary Information

Figure S1 shows the angular magnetoresistance for HoSb at two representative temperatures 2 and 10 K where this compound is in the paramagnetic and antiferromagnetic states respectively. As displayed in Fig. S1(a), the angular magnetoresistance at 10 K simply follows a $B|\cos\theta|$ function (not shown), which is widely observed in GdSb and ErBi[32,30]. In Fig. S1(b), a more complicated angle-dependent magnetoresistance is observed at 2 K. Thus, the spin orderings have significant effect on the angular magnetoresistance of HoSb.

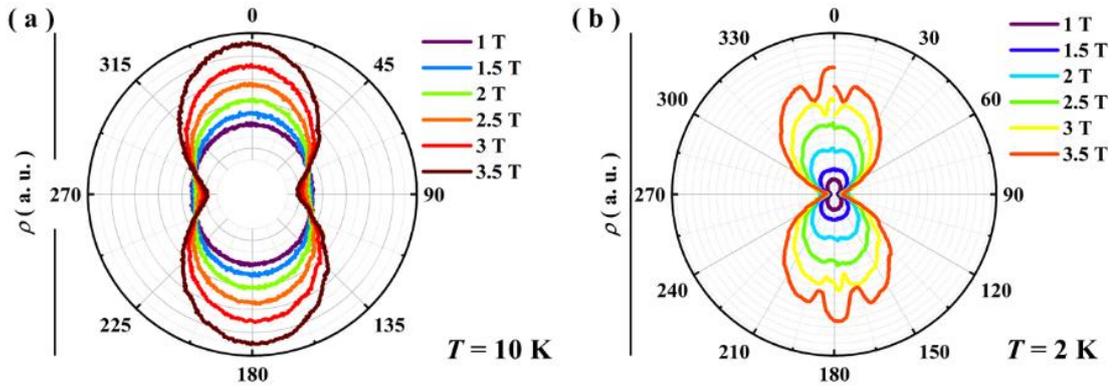

**Figure S1** (a)-(b) The angular magnetoresistance of HoSb single crystal at 10 and 2 K.

To clearly show the topological nature of the electronic properties for NM HoSb in Fig. 5(a), the Wilson loop (Wannier Charge Center) in the $k_z = 0$ plane is calculated by GGA method, from which a nonzero topological invariant can be obtained.

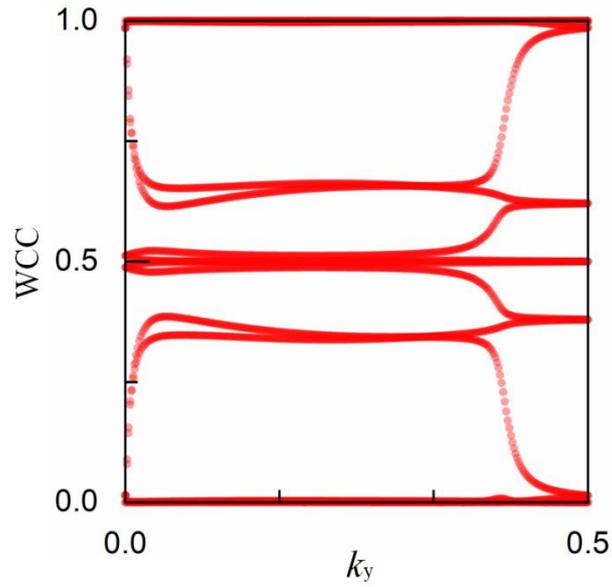

**Figure S2** Wannier charge centers for nonmagnetic HoSb.

**Reference**

[1] J. J. Song, F. Tang, W. Zhou, Y. Fang, H. L. Yu, Z. D. Han, B. Qian, X. F. Jiang, D. H. Wang and Y. W. Du, Extremely large magnetoresistance in the antiferromagnetic semimetal GdSb, *J. Mater. Chem. C*, 2018, **6**, 3026.

[2] L. Y. Fan, F. Tang, W. Z. Meng, W. Zhao, L. Zhang, Z. D. Han, B. Qian, X. F. Jiang, X. M. Zhang and Y. Fang, Anisotropic and extreme magnetoresistance in the magnetic semimetal candidate erbium monobismuthide, *Phys. Rev. B*, 2020, **102**, 104417.